\documentclass[10pt]{iopart}
\usepackage{graphicx}
\usepackage{xcolor}
\usepackage[utf8]{inputenc}
\usepackage{hyperref}
\usepackage{amssymb}

\newcommand{\eV}{\ \mathrm{eV}}
\newcommand{\ket}[1]{|{#1}\rangle}
\newcommand{\braket}[2]{\langle{#1}|{#2}\rangle}
\newcommand{\bra}[1]{\langle{#1}|}
\newcommand{\kv}{\kvec}
\newcommand{\kvec}{\mathbf{k}}

\newcommand{\rv}{\mathbf{r}}
\newcommand{\Rv}{\mathbf{R}}
\def\kv{\kvec}

\newcommand{\eps}{\varepsilon}
\newcommand{\braopket}[3]{\langle{#1}|{#2}|{#3}\rangle}
\newcommand{\idop}{\hat {1\!\!1}}
\newcommand{\svo}{SrVO$_{3}$}

\newcommand{\corrspace}{\mathcal{C}}
\newcommand{\projC}{\hat{P}^{\corrspace}}

\newcommand{\ps}[1]{\tilde{#1}}

\newcommand{\KS}{\Psi_{\nu\kv}}
\newcommand{\KSp}{\Psi_{\nu'\kv}}
\newcommand{\psKS}{\ps{\Psi}_{\nu\kv}}

\newcommand{\ich}{n}
\newcommand{\jch}{n'}

\newcommand{\aew}{\phi}

\newcommand{\psp}{\ps{p}}

\newcommand{\xc}{\mathrm{xc}}
\newcommand{\dc}{\textsc{dc}}

\newcommand{\Sigimp}{\Sigma^{\mathrm{imp}}}
\newcommand{\iomn}{\rmi\omega_{n}}

\newcommand{\VASP}{\textsc{vasp}}
\newcommand{\Wien}{\textsc{wien2k}}
\newcommand{\triqs}{\textsc{triqs}}
\newcommand{\dfttools}{\textsc{dfttools}}
\newcommand{\ks}{\mathtt{KS}}

\newcommand{\dft}{\mathtt{DFT}}
\newcommand{\KSsubset}{\mathcal{W}^C}

\begin{document}
 
 
 \title[Charge self-consistent many-body corrections using optimized PLOs]{Charge self-consistent many-body corrections using optimized projected localized orbitals}
 %
 
 \author{M Sch\"uler$^{1,2}$, O~E Peil$^{3,4,5}$, G J Kraberger$^6$, R Pordzik$^{1,2}$, M Marsman$^7$, G Kresse$^7$, T O Wehling$^{1,2}$ and M Aichhorn$^6$}
 \address{$^1$ Institut f{\"u}r Theoretische Physik, Universit{\"a}t Bremen, Otto-Hahn-Allee 1, 28359 Bremen, Germany}
 \address{$^2$ Bremen Center for Computational Materials Science, Universit{\"a}t Bremen, Am Fallturm 1a, 28359 Bremen, Germany}
 \address{$^3$ Centre de Physique Th\'eorique, \'Ecole Polytechnique, CNRS, Universit\'e Paris-Saclay, 91128 Palaiseau, France}
 \address{$^4$ Department of Quantum Matter Physics, University of Geneva, 24 
 Quai Ernest-Ansermet, 1211 Geneva 4, Switzerland}
 \address{$^5$ Materials Center Leoben Forschung GmbH, A-8700 Leoben, Austria}
 \address{$^6$Institute of Theoretical and Computational Physics,
   Graz University of Technology, NAWI Graz, 8010 Graz, Austria}
 \address{$^7$ Faculty of Physics and Center for Computational Materials Sciences,
 University of Vienna, Sensengasse 8/12, 1090 Wien, Austria}

 \begin{abstract}
 In order for methods combining ab-initio density-functional theory and
 many-body techniques to become routinely used,
 a flexible, fast, and easy-to-use implementation is crucial.
 We present an implementation of a general charge self-consistent
 scheme based on projected localized orbitals in the Projector
 Augmented Wave framework in the Vienna Ab-Initio Simulation
 Package (\VASP{}). We give a detailed description on how the projectors are optimally chosen and how the total energy is calculated. We benchmark our implementation in combination with dynamical mean-field theory: First we study the charge-transfer insulator NiO
 using a Hartree-Fock approach to solve the many-body Hamiltonian. We address the advantages of the optimized against non-optimized projectors and furthermore
 find that charge self-consistency decreases the dependence of the spectral
 function -- especially the gap -- on the double counting. Second, using continuous-time quantum Monte Carlo we
 study a monolayer of SrVO$_3$, where strong orbital polarization
 occurs due to the reduced dimensionality. Using total-energy calculation for structure determination, we find that electronic
 correlations have a non-negligible influence on the position of the apical oxygens, and therefore on the thickness of the single SrVO$_3$ layer.
 \end{abstract}
 
 \maketitle
 \ioptwocol
\section{Introduction}

The advances in the field of nanostructures, where control is now
experimentally possible on an atom-by-atom scale, have given rise to a
strong demand for theoretical simulation tools that are capable of
simulating complex correlated electron systems such as
heterostructures, clusters, or adatom arrays on surfaces. 
There are several successful approaches that can be used for
that. Among the most widely used are 
ab-initio approaches, notably density functional theory (DFT) and
many-body model Hamiltonians.

Modern Kohn-Sham DFT, within the local density approximation (LDA)~\cite{lda_ca} or a
generalized gradient approximation (GGA)~\cite{pbe96}, yields various
ground-state properties including crystal structures quite reliably
for many materials, essentially without any ambiguous input
parameters. 
The auxiliary single-particle energies can be seen as an estimate of
the electronic quasi-particle energies,\cite{goerling-dft-1996}
sometimes giving good qualitative agreement. 
However, (semi)local functionals like LDA or GGA are well known to
fail to capture the physics of strongly-correlated materials such as
Mott insulators, unconventional superconductors, heavy Fermion systems,
or Kondo systems.

This is why, for treating these systems, model Hamiltonians such as
the Hubbard model are usually employed. Generally speaking, these approaches
focus on the description of electron 
correlation phenomena in a minimal low-energy Hilbert space. 
They are accessible by a broad set of many-body techniques including
dynamical mean-field theory (DMFT)\cite{Georges1996} and
generalizations thereof. 
These models naturally depend on model parameters that are
unknown a priori. Keeping the number of parameters low can lead to
over-simplified models that fail to explain sufficiently well
experimental findings. On the other hand, keeping a large number of unknown
parameters makes the results ambiguous already from the outset.

Therefore, to obtain the best of both worlds, it appears natural to
combine the complementary strengths of both approaches to model
correlated electron systems in a realistic manner. 
To achieve this goal, there has been an ongoing effort in the development of
approaches like DFT+DMFT\cite{lichtenstein__1998,Kotliar2006} for more
than 20 years now. 
The class of materials addressed by DFT+DMFT is very wide and includes
Mott insulators, correlated metals, superconductors, and magnetic
materials. 

On the DFT side, simulations using projector augmented wave (PAW)~\cite{paw94}
basis sets turn out to provide a good compromise between accuracy and
computational requirements in these systems. 
Thus, several DFT+DMFT approaches have been implemented with PAW basis
sets on the DFT side.\cite{Amadon2008,karolak_general_2011}
In its most simple formulation, the DMFT is performed on top of a
converged DFT calculation (so-called one-shot DFT+DMFT). 
For many systems, especially those where the strong correlations cause
a rearrangement of charges, the success of the method can be significantly improved by
performing a fully charge self-consistent
calculation.
\cite{Pourovskii2007,haule2009,TRIQS/DFTTools2,bhandary_charge_2016,Granas2012}
There, the electron density obtained from the DMFT calculation is fed back to
the DFT code and the entire loop is self-consistently solved. 
There already exist implementations of DFT+DMFT using the PAW
formalism that include charge self-consistency.
\cite{grieger_approaching_2012,Amadon2012,gonze_recent_2016}

Here we describe a fully charge self-consistent implementation
based on optimized projected local orbitals in the Vienna Ab Initio
Simulation Package (\VASP{}).\cite{paw_vasp,vasp1,vasp2} 
On the DFT side, it is flexible and easy to use; \VASP{} provides the
data necessary to construct the low-energy model of the system and
offers an interface so that the updated charge density can be handed
back. 
This makes it possible to combine it, e.g., with any kind of many-body
correction beyond DFT in some correlated subspace defined via local
projection operators without requiring any changes to the DFT code. 
On the DMFT side, we present two codes making use of this extension of \VASP{}.

The paper is organized as follows.
In \sref{sec:LOCPROJ} we explain the details of our 
approach, particularly, the definition of the optimized local
projectors and the way the charge feedback from the many-body to the
DFT part is implemented. 
We then present an application to the testbed material of NiO in
\sref{sec:NiO}, where we first analyze the optimized versus non-optimized projectors and second demonstrate how full charge
self-consistency affects simulations of the electronic structure,
particularly, in relation with the so-called double-counting problem. 
In \sref{sec:SrVO3}, we report  an application of our 
scheme to monolayers of SrVO$_3$, where we compare our implementation,
in the \triqs{}\cite{triqs} framework, to the already existing interface between
\triqs{} and the DFT code \Wien{}.\cite{Wien2k1,Aichhorn2016}
Furthermore, for that material, we demonstrate total energy
calculations and find structural changes induced by correlations. 

\section{Details of the implementation}
\label{sec:LOCPROJ}

\subsection{Correlated subspace from optimally projected local orbitals}

To perform DFT+DMFT calculations one needs a way to transform between the basis of
the Kohn-Sham (KS) states and the localized basis of the subspace used to define
lattice models, e.g., a Hubbard model. The most obvious way to do this is to
perform a unitary transformation of a subset of KS states to construct a set of
local states. This method is at the heart of various types of Wannier function
based methods, such as the maximally-localized Wannier
functions.\cite{Marzari2012}

In many cases, when KS bands with different characters are strongly entangled,
it becomes however difficult to construct a well-defined unitary
transformation. In this case it is more advantageous to use projector operators
which, acting on the KS Hilbert space, project out only KS states with a
desired character. This is the basis of projected localized orbitals (PLO).\cite{Amadon2008}

In the PLO formalism one starts by defining an orthonormal localized basis set
$\ket{\chi_{L}}$ associated with each correlated site, which is typically
indexed by local quantum numbers, e.g., $L = \{l, m, \sigma,...\}$ with ($l,m$)
and $\sigma$ being orbital and spin angular-momentum quantum-numbers,
respectively. $\{\ket{\chi_{L}}\}$ spans a correlated subspace $\corrspace$ at
each correlated site. Assuming 
$\braket{\chi_{L}}{\chi_{L'}} = \delta_{LL'}$, 
any operator acting on this space can be constructed by
projecting onto $\corrspace{}$ using a projector operator $\projC$, i.e.,
\begin{eqnarray} 
\hat{A}^{\mathrm{imp}} =  \projC \hat{A} \projC, \\ 
\projC =  \sum_{L} \ket{\chi_{L}}\bra{\chi_{L}}.  \label{eq:Pc}
\end{eqnarray} 
An arbitrary vector $\ket{\Psi}$ of the Hilbert space can be decomposed in
terms of local states by writing 
$\projC \ket{\Psi} = \sum_{L} \ket{\chi_{L}} \braket{\chi_{L}}{\Psi}$. 
If we now consider a complete basis
$\ket{\Psi_{\mu}}$, the projector operator is completely defined by specifying
PLO functions $P_{L,\mu} \equiv \braket{\chi_{L}}{\Psi_{\mu}}$.

As described in detail in \cite{Amadon2008,karolak_general_2011}, the PLO
projector in the PAW framework\cite{paw94,paw_vasp} can be written as
\begin{equation}
P^{\Rv}_{L, \nu}(\kv) = \sum_{i}
  \braket{\chi_{L}^{\Rv}}{\aew_{i}} \braket{\psp_{i}}{\psKS}, 
\end{equation}
where $\ket{\chi_{L}^{\Rv}}$ are localized basis functions associated with each
correlated site ${\Rv}$, $\ket{{\aew_{i}}}$ are all-electron partial waves, and
$\ket{\psp_{i}}$ are the standard PAW projectors. In the following, we will
omit the site index $\Rv$ unless it leads to a confusion.
The index $i$ stands for the PAW channel $\ich$, the angular momentum quantum number $l$, and its
magnetic quantum number $m$. $\ket{\psKS}$ are pseudo-KS states. For the PAW potentials distributed
with \VASP{}, generally a minimum of two channels $\ich$ for each angular quantum number are used.
For instance for $3d$ elements, the first $3d$ channel is placed at the energy of the bound $3d$ state
in the atom, and a second channel is added a few eV above the bound atomic $3d$ state. Inclusion of the second
channel improves the transferability of the PAW potentials and the description of the scattering properties greatly. For $s$ and $p$ states in, e.g., $3d$ transition
metals, the first channel usually describes $3s$ and $3p$ semi-core states and the second channel
is placed somewhere in the valence regime ($4s$ and $4p$ states) or even above the vacuum level.
Although a description of how the projectors have been obtained is stored in more recent PAW potential files, 
it is not always straightforward for the user to identify the best suitable PLO projectors. 

Specifically, various choices are possible for
$\ket{\chi_{L}}$.\cite{karolak_general_2011}
Conveniently, 
one might simply use the all-electron partial waves for a particular PAW channel $\ich$,
$\ket{\aew_{L \ich}}$, as the local basis. However, doing so in \VASP{}
can lead to a non-optimal projection. For instance, for transition metals to the left
of the periodic table, the $d$ states in the solid might be more or less contracted than those
in the atom, so that the ``best'' $\ket{\chi_{L}}$ is a linear combination of the 
two available $3d$ channels. Likewise, projection onto the TM valence $s$ states is
often difficult, because of the presence of semi-core $s$ states in the PAW potential file. 
Ideally, the user should not need to make a manual choice of the local $\ket{\chi_{L}}$
functions.

In the following we explain a protocol that largely resolves these issues.
The first step is not strictly required, but simplifies the subsequent coding somewhat.
We first construct a set of PAW projectors and partial waves that are orthogonalized inside the PAW
sphere. This can be achieved by
diagonalizing the all-electron one-center overlap matrix $O_{\ich\jch}$ (for each angular and magnetic quantum number  $L$),
\begin{eqnarray}
O_{\ich\jch} =  \braket{\aew_{L \ich}}{\aew_{L \jch}}, \label{eq:overlap_Q}
\end{eqnarray}
which gives the eigenvalues $\lambda_{\ich}$ and the eigenvector matrix $U$,
\begin{eqnarray}
\Lambda =  \mathrm{diag}(\lambda_{1}, \dots, \lambda_{n}), \\
\Lambda =   U^{\dagger} O U.
\end{eqnarray}
The new partial waves and the corresponding PAW projectors can now be defined as linear combinations
of the original ones:
\begin{eqnarray}
\ket{\xi_{L \ich}} =  \frac{1}{\sqrt{\lambda_{\ich}}} \sum_{\jch}  U_{\ich\jch} \ket{\aew_{L \jch}}, \\
\bra{\tilde{\beta}_{L \ich}} =  \sqrt{\lambda_{\ich}} \sum_{\jch}
   U^{\dagger}_{\jch\ich} \bra{\psp_{L \jch}}.
\end{eqnarray}
This new set of projectors and partial waves have the important 
property that the all-electron partial waves $\ket{\xi_{L \ich}}$ are orthogonal to each other and that they
are normalized to one. Thus, we will now refer to them as ``orthonormal'' partial waves. The choice above is
not unique, though, for instance one might also adopt a Gram-Schmidt orthogonalization procedure.

To construct a unique projector, we seek a unitary transformation that maximizes the overlap between the projector and 
KS valence states \textit{within a chosen (user supplied) 
energy window} $[\eps^P_{\min}, \eps^P_{\max}]$. Specifically, we diagonalize a matrix
\begin{eqnarray}
M_{\ich \jch} =  \sum_{\nu \kv}  \braket{\tilde{\beta}_{L \ich}}{\psKS}
  \braket{\psKS}{\tilde{\beta}_{L \jch}},
\end{eqnarray} 
where the sum is restricted to states with energies $\eps_{\nu \kv}\in[\eps^P_{\min}, \eps^P_{\max}]$. We then pick the eigenvector $\upsilon_{\ich}$ with the largest absolute eigenvalue to construct
local basis functions and corresponding PAW projectors,
\begin{eqnarray}
\ket{\chi_{L}} =  \sum_{\ich} \upsilon_{\ich} \ket{\xi_{L \ich}}, \\
\bra{\tilde{\pi}_{L}} =  \sum_{\ich} \upsilon^{*}_{\ich} \bra{\tilde{\beta}_{L \ich}},
\end{eqnarray}
with projector functions given by\footnote{The projectors $P_{L,\nu}(\kv)$ according to \eref{eq:projector-functions} containing all phase factors are written by \VASP{} to a file called LOCPROJ.}
\begin{eqnarray}
P_{L,\nu}(\kv) = \braket{\tilde{\pi}_{L}}{\psKS}.
\label{eq:projector-functions}
\end{eqnarray}
We note in passing that both steps could be combined into a single computational step, by
diagonalization of a generalized eigenvalue problem involving the overlap matrix $O$ and the matrix $M$ evaluated using 
the original set of projectors and partial waves. 

Note that since the all-electron partial waves do not form an orthonormal basis set between different PAW spheres and
because projection is performed for a subset of KS states,
the above procedure will usually produce non-normalized local states.
However, normalized local states can be constructed if we orthonormalize the projector
functions as follows,
\begin{eqnarray}
O_{L L'} =  \sum_{\nu\kv} P_{L,\nu}(\kv) P^{*}_{\nu, L'}(\kv), \label{eq:projector_overlap} \\
P_{L,\nu}(\kv) \leftarrow  \sum_{L'} O_{L L'}^{-\frac{1}{2}} P_{L',\nu}(\kv). \label{eq:projector-orthonormalized} 
\end{eqnarray}

On a technical level, the optimal projectors are constructed internally in \VASP{} 
according to (\ref{eq:overlap_Q}-\ref{eq:projector-functions})
for each sites $\Rv$ given an energy window $[\eps^P_{\min}, \eps^P_{\max}]$.
The last orthonormalization step \eref{eq:projector-orthonormalized}
is done externally as post-processing. This allows one to choose a different
energy window for the summation in \eref{eq:projector_overlap} in order
to fine-tune the degree of localization of the impurity Wannier functions.

The projection scheme has been used quite widely in the solid state community,\cite{Amadon2008,karolak_general_2011} even 
in combination with VASP. When using VASP, these calculations were often not based
on a solid mathematical foundation. VASP used to project onto all available PAW projectors
with a certain angular and momentum quantum number $lm$, summed the resulting densities, averaged the phase factors and intensities over all $lm$ projectors,
and finally wrote the results to a file ({\tt PROCAR}). Such a prescription, in particular, the averaging
of the intensities and phase factors was done {\em ad hoc} without a proper mathematical prescription. This
can  result in unsatisfactory results if the two projectors span a completely different subspace, for
instance Ni $3d$ and  $4d$ states, or transition metal semi-core $p$ states and valence $p$ states. 
We will demonstrate this issue for NiO in \sref{sec:NiO_proj}.

\subsection{Total energy and charge self-consistency}

A general extension of DFT Kohn-Sham equations to a many-body problem based on the Hubbard model can
be formulated using the total-energy functional\cite{Savrasov2004,Pourovskii2007}
\begin{eqnarray}
\eqalign{
 E  [n,& {\hat G}^\mathcal{C} ] = \frac{1}{\beta} \sum_{\iomn\kv} \Tr\{  \hat G(\kv,\iomn) \hat H_{\ks}(\kv)\}   \\
 &  +E_{H}[n] + E_{\xc}[n]  - \int d\rv\, (v_{\xc} + v_{H}) n(\rv)  \\
 &  +E_{\mathrm{corr}}[{\hat G}^\mathcal{C}] - E_{\dc}[{\hat G}^\mathcal{C}],  \label{eq:etot_dmft} 
 }
\end{eqnarray}
where ${\hat G}(\kv,\iomn)$ is the Green's function containing many-body effects, and $ \hat H_{\ks}(\kv)=\sum_{\nu} \ket{\Psi_{\nu\kv}} \varepsilon_{\nu \kv} \bra{\Psi_{\nu\kv}}$ is the KS Hamiltonian.
The charge density, $n(\rv)$, and Green's function in the correlated subspace, ${\hat G}^\mathcal{C}(\kv,\iomn)$, are
both related to $ \hat G(\kv,\iomn)$ by
\begin{eqnarray}
n(\rv) =  \frac{1}{\beta} \Tr \sum_{\iomn\kv} \braopket{\rv}{\hat G(\kv,\iomn)}{\rv}, \\
{\hat G}^\mathcal{C}(\kv,\iomn) =  \projC(\kv)  \hat G(\kv,\iomn) \projC(\kv),
\end{eqnarray}
where $\projC_\kv$ projects onto correlated localized states, see \eref{eq:Pc}.
$E_{\mathrm{corr}}$ is the energy contribution of the interaction term
(Hubbard $U$-term) and $E_{\dc}$ is the double-counting correction.

The first four terms of the above expression form the usual DFT total energy
calculated for the density matrix and charge density 
of the  {\em interacting} system, which is non-diagonal in this case.
It is thus clear that to get the correct value of the
total energy the density matrix must be calculated in a self-consistent manner.

In the framework of DFT+DMFT \cite{Georges1996,Kotliar2006} the interacting Green's function is defined
as follows:
\begin{eqnarray}
\hat G(\kv,\iomn) =\left[(\iomn + \mu) \idop -  \hat H_{\ks}(\kv) - 
\hat \Sigma^{\ks}(\kv, \iomn) \right]^{-1}\hspace{-0.35cm},
\end{eqnarray}
where  $\hat \Sigma^{\ks}(\kv, \iomn)$ is obtained by up-folding the local self-energy,
\begin{eqnarray}
\eqalign{
\hat \Sigma^{\ks}(\kv, \iomn) = &\sum_{\nu \nu'}   \ket{\KS}\bra{\KSp}
	 \\
	&\cdot \sum_{m m'} P^{*}_{\nu,m}(\kv) \Sigma_{mm'}(\iomn) P_{m',\nu'}(\kv) ,
}
\end{eqnarray}
where the local self energy's matrix elements in the basis of localized impurity states, $\ket{\chi_{lm}}$, are
\begin{eqnarray}
 \Sigma_{mm'}(\iomn) =  \Sigimp_{mm'}(\iomn) - \Sigma^{\dc}_{mm'},
 \label{eq:self_energy_dc}
\end{eqnarray}
which consist of the purely local impurity self-energy $\hat \Sigma^{\mathrm{imp}}(\iomn) $, obtained from an impurity solver, and the double counting $\hat \Sigma^{\dc}$. Using the PLO functions we can write this Green's function in terms of its
KS matrix elements
\begin{eqnarray}
\eqalign{
G_{\nu\nu'}(\kv, \iomn) =  \Big [ (\iomn +&  \mu -  \eps_{\nu\kv}) \idop \\
& - \hat \Sigma^{\ks}(\kv, \iomn)\Big]^{-1}_{\nu\nu'}.
}
\end{eqnarray}

If we define an energy window $[\eps^C_{\min}, \eps^C_{\max}]$ which selects a subset
$\KSsubset$ of KS states affected by correlations, the interacting charge density can be written as
\begin{eqnarray}
\eqalign{
n(\rv)& =  \sum_{\kv} \sum_{\nu \notin \KSsubset} f_{\nu\kv} \braket{\rv}{\KS} \braket{\KS}{\rv}  \\
 &+ \sum_{\kv} \sum_{\nu\nu' \in \KSsubset}  \braket{\rv}{\KS} N_{\nu\nu'}(\kv) \braket{\KSp}{\rv}, 
\label{eq:chden}
}
\end{eqnarray}
where we use a non-diagonal density matrix
\begin{eqnarray}
N_{\nu\nu'}(\kv) = & \frac{1}{\beta} \sum_{\iomn} G_{\nu\nu'}(\kv, \iomn),
\end{eqnarray}
and a diagonal density matrix formed by the KS occupation numbers $f_{\nu\kv}$.
Note that the chemical potential $\mu$ here is generally different from the KS
chemical potential $\mu_{\ks}$. The subset of KS states for the optimization of the projectors $\mathcal{W}^P$ and KS states affected by correlations $\KSsubset$ can be chosen to be the same but one does necessarily not have to do so.




From a practical point of view it is convenient to split the charge density into
a DFT part and a correlation-induced part,
\begin{eqnarray}
n(\rv) = & n_{\dft}(\rv) + \Delta n(\rv), \\
n_\dft(\rv) = & \sum_{\nu\kv} f_{\nu\kv} \braket{\rv}{\KS} \braket{\KS}{\rv}
\end{eqnarray}
where $\Delta n(\rv)$ is the correlation correction,
%
\begin{eqnarray}
\Delta n(\rv) =  
\sum_{ \stackrel{\kv}{ \nu\nu' \in \KSsubset} }\braket{\rv}{\KS}\Delta N_{\nu\nu'}(\kv) \braket{\KSp}{\rv}, \\
\Delta N_{\nu\nu'}(\kv) =  N_{\nu\nu'}(\kv) - f_{\nu\kv} \delta_{\nu\nu'},
\end{eqnarray}
with the last quantity having the important property
\begin{eqnarray}
  \sum_{\nu \in \KSsubset} \sum_{\kv} \Delta N_{\nu\nu}(\kv) = 0.
\end{eqnarray}
Such a definition of the new charge density is convenient because it
ensures charge neutrality between DFT iterations.


In a similar way, the total energy can be split into a DFT part calculated using the
correlated charge density given by \eref{eq:chden} and a correlation part,
\begin{eqnarray}
\eqalign{
E =  E_{\dft}[n] + 
& \sum_{\kv} \sum_{\nu \in \KSsubset} \Delta N_{\nu\nu}(\kv) \eps_{\nu\kv}  \\
&+ E_{\mathrm{corr}}[\hat G^\corrspace] - E_{\dc}[\hat G^\corrspace]. \label{eq:vasp_toten}
}
\end{eqnarray}

Thus, in order to calculate the total energy and to obtain the charge density for the next
KS iteration, only two quantities need to be calculated after a DMFT iteration: the
density-matrix correction $\Delta N_{\nu\nu}(\kv)$ and the interaction energy (including the
double-counting term) $E_{\mathrm{corr}} - E_{\dc}$.

In this particular \VASP{} implementation, we use $\Delta N_{\nu\nu}(\kv)$ to obtain the natural orbitals 
by a transformation $V$ given by diagonalizing the total correlated
density matrix,
\begin{eqnarray}
f'_{\nu\kv} \delta_{\nu\nu'} =  \sum_{\mu\mu'}V_{\nu \mu} \left[f_{\mu\kv} \delta_{\mu\mu'} \hspace{-0.1cm} + \hspace{-0.1cm}
  \Delta N_{\mu\mu'}(\kv)\right] V^*_{\mu'\nu'}, \\
\ket{\KS'} = \sum_{\mu} V_{\nu\mu} \ket{\Psi_{\mu\kv}},
\end{eqnarray}
and the charge density and the one-electron energy are, then, obtained in the same
way as in the normal KS cycle,
\begin{eqnarray}
n(\rv) =  \sum_{\nu\kv} f'_{\nu\kv} \braket{\rv}{\KS'} \braket{\KS'}{\rv}, \\
\eqalign{
\frac{1}{\beta} \sum_{\iomn\kv} \Tr\{G(\kv, \iomn)& H_{\ks}(\kv)\} =\\
&
  \sum_{\nu\kv} f'_{\nu\kv} \braopket{\KS'}{H_{\ks}}{\KS'}. \label{eq:tr_gh}
  }
\end{eqnarray}

Note that the DFT band energy is calculated using the original occupancies $f_{\nu\kv}$, and
the second term in \eref{eq:vasp_toten} represents essentially a band-energy correction
which takes into account the change in the density matrix induced by correlations.
The occupancies $f'_{\nu\kv}$ will deviate from the usual Fermi-Dirac statistics
as a result of particle fluctuations from the occupied non-interacting KS states into unoccupied
states.

\section{Benchmark for NiO}
\label{sec:NiO}

NiO is a prototypical charge-transfer insulator where the electronic
band gap on the order of $4\eV$ arises from a combination of strong local Coulomb
repulsion $U$ within the Ni $3d$ shell and the charge-transfer energy
$\Delta=\epsilon_d-\epsilon_p$, i.e., the difference in on-site
energies of O-$2p$ and Ni-$3d$ orbitals.
\cite{sawatzky_magnitude_1984} The uppermost valence states are of
hybrid Ni-$3d$ and O-$2p$ character, where the O-$2p$ contribution is
dominant. 

DFT in (semi)local approximations like LDA or
GGA fails to describe the insulating nature of NiO. Non-spin-polarized
LDA and GGA yield a metal, while spin-polarized calculations of the
antiferromagnetically ordered phase of NiO yield an energy gap of
$\sim0.7\eV$, which is much smaller than the experimentally
established gap.\cite{bengone_implementation_2000} The reason for
this failure of semilocal DFT to describe the insulating state of NiO
is well known to be the insufficient treatment of the strong local
Coulomb interaction in the Ni $3d$ shell. NiO has thus become a
testbed material for realistic correlated electron approaches such as
DFT+U \cite{anisimov_band_1991} or DFT+DMFT.\cite{lichtenstein__1998,Vollhardt_2006,KAROLAK201011,PhysRevB.94.155135}
\subsection{Optimized projectors}
\label{sec:NiO_proj}
\begin{figure}
        \centering
        \includegraphics[width=\linewidth]{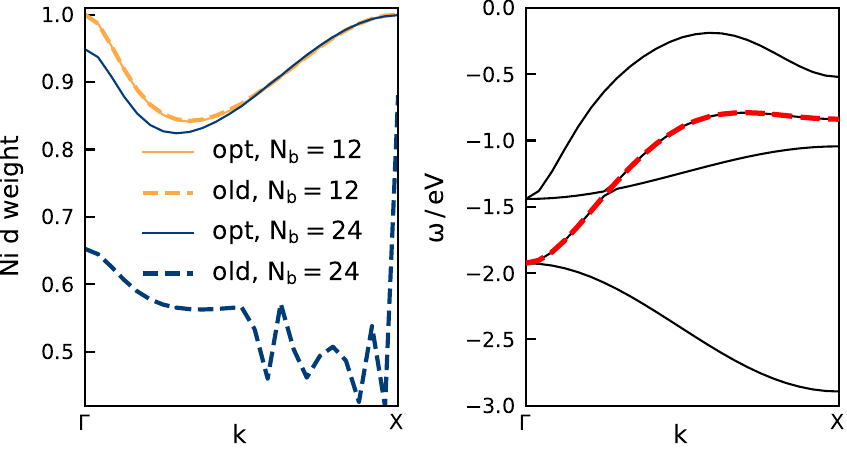}
        \caption{Left: Ni $d$ weight of the band shown as red dashed line on the right for paramagnetic NiO. Results using non-optimized projectors are shown as dashed lines, those from the optimized projectors as solid lines. Orange and blue are results for 12 and 24 bands, respectively.}
        \label{fig:proj_compare}
\end{figure}

For the example of NiO, we first investigate the properties of the optimized projectors described in \sref{sec:LOCPROJ}. To this end, we analyze the Kohn-Sham Hamiltonian transformed to a localized basis, which will later serve as a starting point for including local correlation effects on a Hartree-Fock level.

In order to treat the Ni $d$ states and ligand O $p$-states explicitly, we project the paramagnetic Kohn-Sham Hamiltonian of a two atomic unit cell using $48\times 48 \times 48$  $k$ points onto Ni $d$ and O $p$ states according to
\eref{eq:projector-functions}. Therefor, we optimize the projectors in an energy window around the Fermi level given by $\eps^P_{\min}=-3.3$\,eV
and $\eps^P_{\max}=1.7$\,eV  (i.e., we mainly optimize the Ni $d$ states) and orthonormalize according to
\eref{eq:projector-orthonormalized} for all available energies. We perform a Gram-Schmidt procedure to orthogonally complement the Ni $d$ and O $p$ states with additional states to end up with the same number of localized states as Kohn-Sham states.

In \fref{fig:proj_compare} we analyze the orbital properties of selected eigenstates of the resulting Hamiltonian. The left panel shows the Ni $d$ weight of a single band (displayed as red dashed line in the right panel) across a k-path from the $\Gamma$ to the $X$ point. For a small number of unoccupied bands, (total number of bands $N_b=12$), the non-optimized and optimized projectors give nearly identical results. However, by including additional unoccupied bands ($N_b=24$) the two schemes display huge differences. While the optimized projectors lead to only minor differences to the case of $N_b=12$  close to the $\Gamma$ point, the non-optimized projectors lead to considerably less overall Ni $d$ weight and a strongly discontinuous behavior for part of the $k$ pathway. This underlines the advantage of the optimized projectors.

The erroneous behavior of the non-optimized projectors roots in additional high energy bands having Ni-$4d$ character.
Using the standard VASP ``projectors'', these high lying valence orbitals show appreciable overlap with the $4d$ partial waves (and $4d$ projectors). 
VASP used to lump the $4d$ and $3d$ contributions into single values, which causes the issues we have just observed. Without going into mathematical
details one can easily understand this behavior.  The $3d$ or $4d$ states are automatically orthogonal, because the $4d$ states possess an additional 
node inside the PAW sphere. 
By simply adding up the contributions from the $3d$ and $4d$ projectors pretending that they correspond to a single main quantum number, and then performing
an orthogonalization, the total charge in each $d$ spin channel is one instead of two, explaining the 
large reduction of the $d$ character using the old scheme.

The failure of the unoptimized projectors could in principle be avoided by
simply not including additional empty bands in the construction of the
Wannier functions. However, there are important cases where this is
not possible. First of all, the situation
of $4d$ states close to the Fermi energy is realized in many late
transition metals. Second, one might be interested in quantities that
include transitions to unoccupied states over large energy scales,
such as optical spectroscopy. In these cases, the optimized
projectors give reliable results. 
Furthermore, from a physical point of view, the
projectors should not depend strongly on the inclusion of unoccupied
states.

\subsection{Hartree-Fock Approximation}
The DFT+U approach improves the LDA and GGA description of NiO by supplementing the Ni $d$ states by local Coulomb interaction which leads to a static local self energy in \eref{eq:self_energy_dc}.

There are two ways to formulate DFT+U: First, the traditional one,
where the Kohn-Sham potential is directly augmented with a potential
arising from the Hartree-Fock decoupling of the local Hubbard
interaction. Second, in terms of a charge self-consistent DFT+DMFT
scheme, where the DMFT impurity problem is solved in the Hartree-Fock
approximation called DFT+DMFT(HF) in the following. Both schemes are fully equivalent if the augmentation spaces are
chosen to be strictly the same and the way spin-polarization is
accounted for is the same. 

However, many DFT+U implementations, including the one in \VASP{}, work
with angular-momentum-decomposed charge densities, which in general do
not relate to proper projectors in the definitions of the "+U"
potentials.\cite{Amadon_JPCM_2012}
Additionally, DFT+DMFT and DFT+U for magnetic
materials can work with or without spin polarization in the DFT
part.  

In the following, we compare DFT+U as implemented in \VASP{} to DFT+DMFT(HF) with and
without charge self-consistency for the testbed material NiO, and investigate the dependence of the electronic spectra on the double-counting. 

  \begin{figure}
    \centering
    \includegraphics[width=\linewidth]{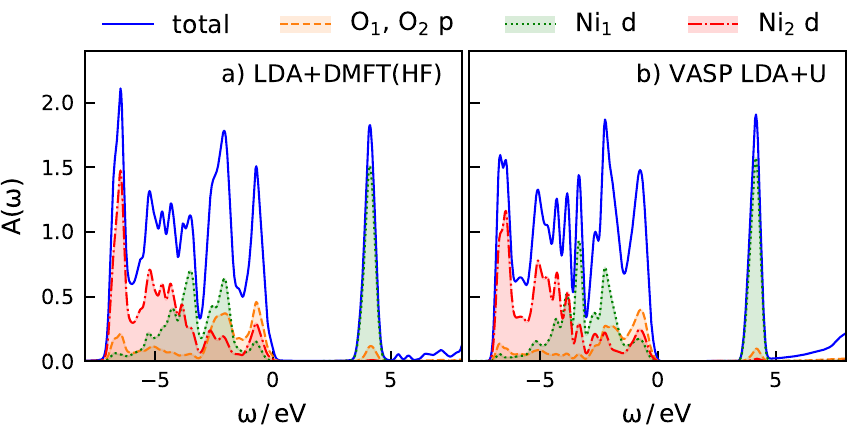}
    \caption{Partial density of states for spin up of Ni $d$ and O $p$
      states from a) charge self-consistent DFT+DMFT(HF) with $\mu^\dc=62.5\eV$ and b) \VASP{}
      DFT+U with FLL double counting. 
        Ni$_1$ and Ni$_2$ refers to the two different Ni atoms in the
        unit cell arising due to the anti-ferromagnetic
        ordering. The spin-up density of the atom Ni$_1$ is equal to
        the spin-down density of Ni$_2$ and vice versa. 
        The oxygen atoms in the unit cell show no spin polarization,
        thus, O$_1$ is equivalent to O$_2$.
    }
    \label{fig:nio_fll}
    \end{figure}

We fix the double counting in the \VASP{} DFT+U approach according to a prescription known as ``fully localized limit'' (FLL)\cite{anisimov_density-functional_1993} which leads to an orbital independent and diagonal version of $\Sigma_{mm'}^\dc$ in \eref{eq:self_energy_dc}, in short $\mu^{\dc}$. 
For a meaningful comparison of spectral features we choose the double counting potential in the DFT+DMFT(HF) approach such that we reproduce the size of the gap in \VASP{} DFT+U, leading to $\mu^\dc = 62.5\eV$. 

We sample the Brillouin zone of the unit cell of twice the size of the primitive one using $8\times8\times8$ k-points and parametrize the Coulomb interaction using Slater integrals given by effective Coulomb parameters $U=8$\,eV and $J=1$\,eV as obtained from constrained LDA calculations.\cite{anisimov_band_1991} We perform spin polarized calculations considering antiferromagnetic ordering of the two Ni atoms in the unit cell, where for both, DFT+DMFT(HF) and \VASP{} DFT+U, we only consider spin polarization in the Hartree-Fock part but not in the DFT part. 

The respective density of states for spin up are presented in \fref{fig:nio_fll}. In general, the
DOS from DFT+U and DFT+DMFT(HF) are very similar:  a gap of about $4$\,eV is
opened between heavily spin-polarized Ni states in the conduction band
and strongly hybridized Ni-O states in the valence band. Only details
differ for the two approaches, such as a slightly larger spin
polarization of Ni states and a larger O contribution at the valence
band edge for DFT+DMFT(HF).

      \begin{figure}
        \centering
        \includegraphics[width=\linewidth]{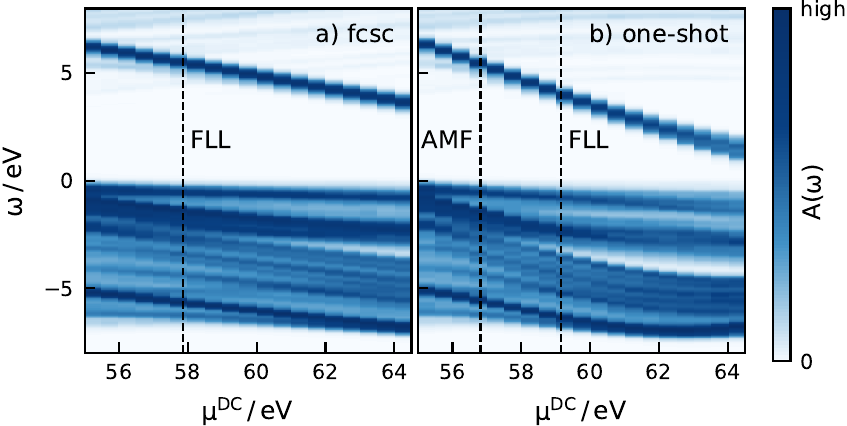}
        \caption{Color-coded density of states from DFT+DMFT(HF) for
          double-counting potentials $\mu^\dc$ in steps of $0.5$\,eV
          for a) full charge self-consistency and b) one-shot
          calculations. The respective $\mu^\dc$ from FLL and AMF are
          depicted as dashed lines, where AMF in the case of charge
          self-consistency is $53.5$\,eV and therefore out of the
          plotting range.}
        \label{fig:nio_map}
        \end{figure}

The double-counting problem poses a serious problem when using
many-body corrections in DFT+DMFT approaches predictively, since
fundamental properties such as the single particle gap depend on the
double counting. Here, we perform fully charge self-consistent (fcsc)
as well as one-shot DFT+DMFT(HF) calculations for various fixed
double-counting potentials $\mu^\dc$ to investigate its influence on
the spectra and on the gap.

Strictly speaking, this
approach does not represent a proper double-counting scheme on first
sight, since there seems to be no functional relation between the
double-counting potential $\mu^\dc$ and energy $E_\dc$. However, one
can motivate this approach following arguments in
\cite{millis_dc}. Instead of using the standard FLL form for the
double-counting correction, one introduces an interaction parameter
$U'$ in the double counting formulas, which is allowed to be slightly
different from $U$ used otherwise in the calculation. In that way, one
can again relate $\mu^\dc$ to a proper energy 
correction $E_\dc$.

However, here we are not interested in
total energies but in a systematic
investigation of the influence of the double counting on the spectral
properties. That is why we refrain from an explicit introduction of
this parameter $U'$. Furthermore, we want to circumvent here
further complicating ambiguities in usual double-counting approaches
(e.g., if one should use the formal or self-consistent occupation in
formulas for $E_\dc$).\cite{KAROLAK201011}

The resulting total DOS for the fcsc and one-shot
calculations are presented color coded in \fref{fig:nio_map}.  
In both cases the Ni states in the conduction band (dark blue around
$5$\,eV) are shifted linearly towards the Fermi energy as $\mu^{\dc}$ increases. 
The O/Ni states at the valence band edge are not affected strongly by
the double counting. 
This is because, in general, the higher the Ni character of a band,
the more its energy is shifted by the double-counting correction.

In the conduction band we also find states with low spectral weight in the PAW spheres
(i.e., they are neither located around the Ni nor the O atoms but in the interstitial region); their energies shift to lower values with decreasing
$\mu^{\dc}$, i.e., in the opposite direction than the 
Ni state in the conduction band. 
For $\mu^\dc \lesssim 58$\,eV in case of fcsc and
$\mu^\dc \lesssim 57$\,eV  in the case of one-shot calculations, these states cross the Ni
states. Then, the conduction band edge
does not consist of Ni states, which is not in agreement with
experiment. 
Note that the ``around-mean-field'' AMF prescription lies in this
regime both for one-shot 
and fcsc calculations and is thus not suitable for NiO. Similarly, the FLL prescription for fcsc calculations is very close to this regime.

For values of the double counting that give the correct conduction
band character, the gap shrinks with increasing $\mu^\dc$. 
This dependence differs strongly for the one-shot and charge
self-consistent calculations: the slope of the gap $d E_g / d\mu^\dc$
in the physical regime differs by a factor of nearly 2. 
Thus, the charge self-consistency does not cure but alleviates the
double-counting problem by decreasing the influence of $\mu^\dc$ on
the spectrum. 

\section{Benchmark for SrVO$_3$ monolayer}
\label{sec:SrVO3}

\svo{} presents an example of a correlated transition-metal oxide experiencing a
metal-insulator transition (MIT) driven by a dimensional crossover.~\cite{Yoshimatsu2010}
In particular, a monolayer of \svo{} grown on a SrTiO$_{3}$ substrate is an insulator
with a Mott gap of around 2\,eV. In DFT, the compound is found to be metallic and one has to
combine DFT with a many-body technique to achieve the correct insulating solution. The related problem of a double layer of \svo{} shows insulating behavior in a non charge self-consistent DFT+DMFT treatment. \cite{zhong_electronics_2015}

Here, we use a monolayer of \svo{} to benchmark the fully charge-self-consistent (fcsc)
DFT+DMFT implementation of \triqs{}/\allowbreak\dfttools{}\cite{Aichhorn2016} within the \VASP{} PLO formalism.
The motivation for choosing this particular system is that
electronic correlations induce an appreciable charge redistribution,
making the use of a fcsc DFT+DMFT scheme imperative.
Importantly, using the \triqs{}/\allowbreak\dfttools{} framework, we can benchmark the 
presented \VASP{} interface to the one that
is based on the \Wien{} DFT package,\cite{Wien2k1} also implemented in
\triqs{}/\dfttools{}. Furthermore, we
compare our results to previously published fcsc data, also based on
the \Wien{} code, but using MLWF as correlated basis set.\cite{bhandary_charge_2016}


The system is modeled by a free-standing monolayer of \svo{} with the in-plane
lattice constant equal to that of SrTiO$_{3}$ (3.92\,\AA{}), simulating thus the
epitaxial geometry. The effect of the substrate is neglected. To isolate the individual
layers in a periodic unit cell, a vacuum layer of about 16\,\AA{} is used in the \VASP{} calculations.

Based on geometry relaxations on the DFT level, in the out-of-plane
direction, the V-O distance is reduced from $1.96$\,\AA{} to $1.93$\,\AA{} and the Sr-Sr
distance from $3.92$\,\AA{} to $3.52$\,\AA{}. 
The Brillouin zone is sampled using a $15\times 15\times 1$
$\Gamma$-centered Monkhorst-Pack $k$-grid. 
The energy cutoff of the plane wave basis set is $400$\,eV,
in accordance with the default value for the PAW potentials
used. 
Using the procedure described above, the projectors onto the V
$d$-states are calculated according to
\eref{eq:projector-functions} by optimizing in an energy window around the Fermi level given by $\eps^P_{\min}=-2$\,eV
and $\eps^P_{\max}=1.1$\,eV (i.e., states with mainly $t_{2g}$ character) and orthonormalizing according to
\eref{eq:projector-orthonormalized} in the same energy window. 
In the $t_{2g}$ subspace, a Hubbard-Kanamori interaction with $U =
5.5$\,eV and $J = 0.75$\,eV (in agreement with
\cite{bhandary_charge_2016}) is added; the resulting
double counting is estimated in the FLL scheme.\cite{anisimov_density-functional_1993,held_dc}
The impurity problem is solved using the \triqs{}/\textsc{CTHYB} Quantum Monte 
Carlo\cite{Seth2016} solver at an inverse temperature $\beta = 40\ 
\mathrm{eV}^{-1}$. 

\begin{figure}
    \centering
    \includegraphics[width=0.6\linewidth]{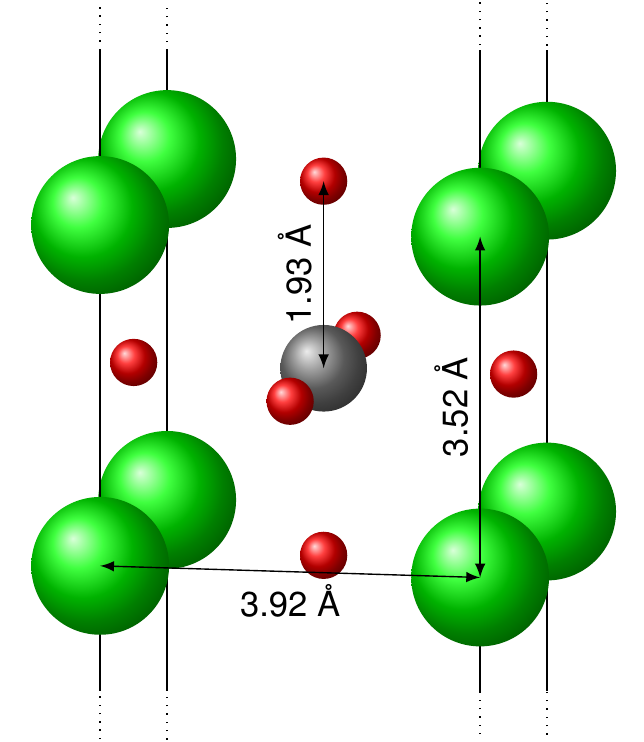}
    \caption{The unit cell of the monolayer of SrVO$_3$.
    The Sr atoms are green, the V atom gray and the O atoms red.
    The lattice constant in the $z$-direction is $20$\,\AA{}, i.e.,
there is about 16\,\AA{} of vacuum between the periodic replica of the layers, effectively giving an isolated monolayer.}
    \label{fig:srvo3-uc}
\end{figure}

\begin{figure}
    \centering
    \includegraphics[width=\linewidth]{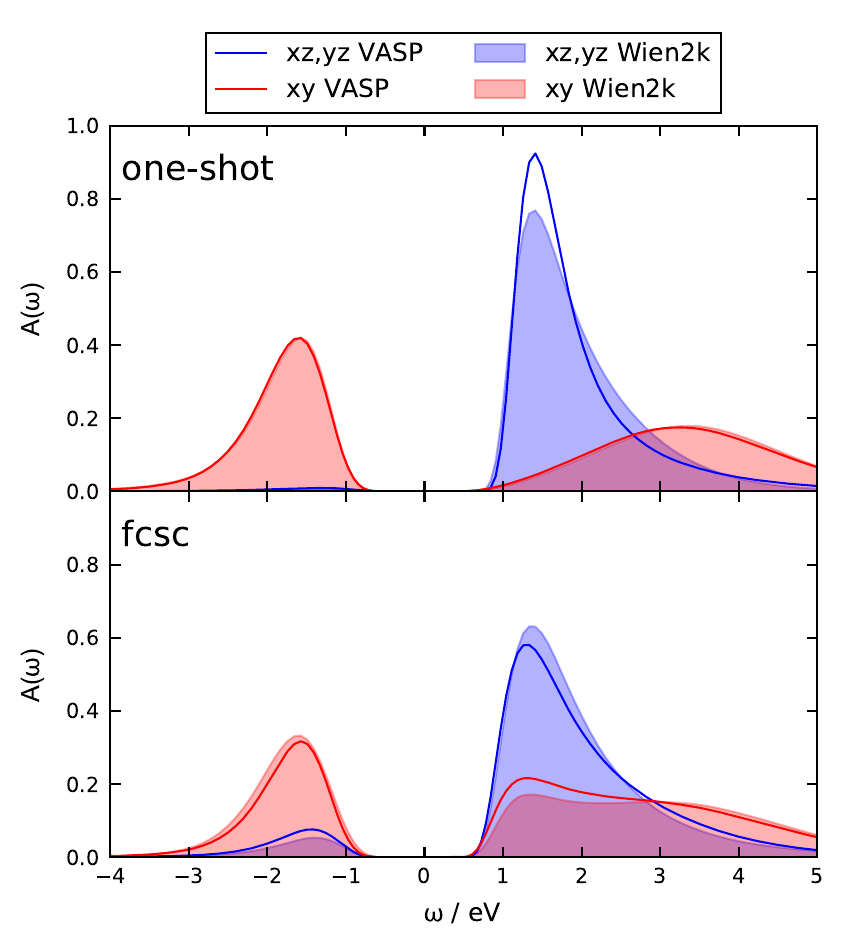}
    \caption{DFT+DMFT spectral function of the single layer of SrVO$_3$ in a one-shot (top) and 
        fully charge self-consistent calculation (bottom).
        The calculations have been performed using \triqs{}/\dfttools{},
        once with \Wien{} (compare \cite{bhandary_charge_2016}) and once with \VASP{} as underlying DFT code.
        The resulting imaginary-time Green's function was analytically
        continued using the Maximum Entropy Method.\cite{bryan_1990}} 
    \label{fig:srvo3-spectral}
\end{figure}

The one-shot DFT+DMFT calculation results in a nearly complete polarization of the orbitals
(see fillings in \tref{tab:srvo3-filling}), which is found to be
equal in \VASP{} and \Wien{}, and which is also in accordance with
published data.\cite{bhandary_charge_2016}
When using the charge feedback in the fcsc framework, the empty orbitals become partly repopulated.
This effect happens slightly stronger in \VASP{} but the agreement
of the two calculations based on the two different DFT codes is within
the expected difference between the two implementations.
This re-population can also be seen in the spectral function (\fref{fig:srvo3-spectral}), 
which is obtained using analytic continuation of the local lattice
Green's function using the maximum entropy method.\cite{bryan_1990}
Unlike the one-shot calculation (top panel of \fref{fig:srvo3-spectral}), 
the fcsc scheme produces a lower 
Hubbard band with non-negligible spectral weight 
also for the degenerate $d_{xz}$ and $d_{yz}$ orbitals (bottom panel of \fref{fig:srvo3-spectral}).
The same Mott gap (whose value is in good agreement with
  experiment \cite{Yoshimatsu2010}) is found starting from both DFT
codes,  
with the peak positions being basically identical.
The difference in peak height is compatible with the difference of
filling between the two methodologies. The calculated spectra are in
excellent agreement with results presented in \cite{bhandary_charge_2016}.

\begin{table}
\caption{\label{tab:srvo3-filling} Filling of the correlated orbitals in DFT+DMFT for one-shot and fully charge self-consistent calculations based on \VASP{} and \Wien{}.}
\begin{indented}
\centering
\item[]\begin{tabular}{@{}lcccc}
\br
 & \multicolumn{2}{c}{one-shot} & \multicolumn{2}{c}{fcsc} \\
 
 & $d_{xy}$ & $d_{xz}$, $d_{yz}$ &  $d_{xy}$ & $d_{xz}$, $d_{yz}$ \\
 \mr
\VASP{}  & 0.96 & 0.02 & 0.68 & 0.16\\
\Wien{} & 0.96 & 0.02 & 0.76 & 0.12\\
\br
\end{tabular}
\end{indented}
\end{table}

\begin{figure}
    \centering
    \includegraphics[width=\linewidth]{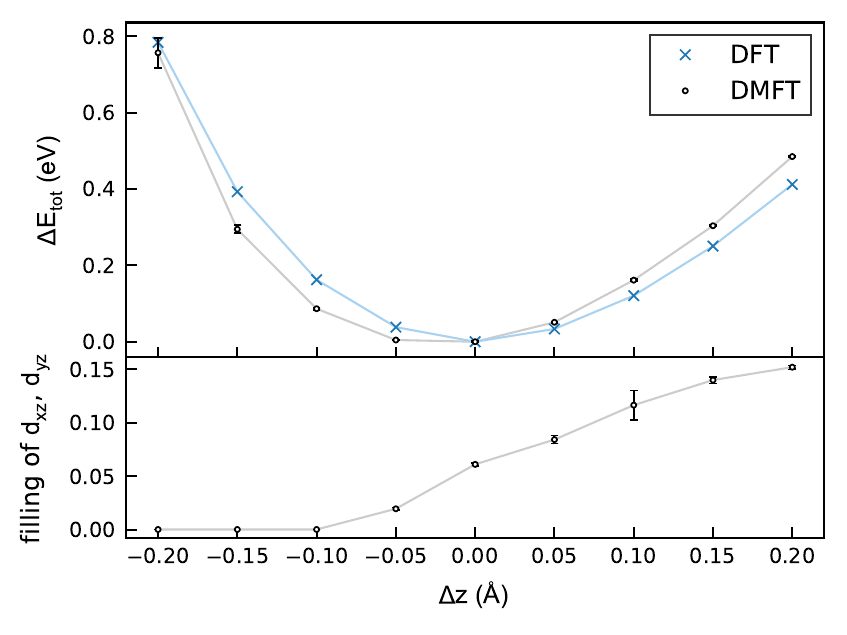}
    \caption{
        Top: Total-energy change of the single layer of SrVO$_3$ when
        moving the upper and lower Sr-O-plane symmetrically  
        by $\Delta z$ (from the DFT-optimized structure for $\Delta z=0$) in DFT and in fully charge self-consistent
        DFT+DMFT. 
        For convenience, the energy for $\Delta z = 0$ is shifted to
        $0$ in each case. 
        Bottom: Filling of the degenerate $d_{xz}$ and $d_{yz}$ orbitals per spin channel as a function of $\Delta z$.
        The total filling of the impurity is 1 electron.
        The lines in both plots are guides to the eye.
        The values and error bars of the DFT+DMFT calculations in both plots were obtained by calculating the quantity for 
        the four last iterations and then determining the mean and the standard deviation.
        For many data points the error bars are smaller than the markers.
        The total energy in DFT was converged to $10^{-6}$\,eV.
    }
    \label{fig:svo-toten}
\end{figure}

Full charge self-consistency allows us to calculate reliably the total energy as a
function of structural parameters and to determine the lowest-energy structure in DFT+DMFT.
Here, as a proof of principle, we calculate the total energy of the compound as a function of
the distance between Sr-O and V-O planes.
We consider deviations $\Delta z$ from the DFT-optimized structure.
Positive $\Delta z$ means that the upper Sr-O plane gets shifted upwards and the lower plane gets shifted downwards, 
thus increasing the thickness of the slab in $z$ direction.
This changes the splitting between the $d_{xy}$, which is close to half-filling, 
and the degenerate $d_{xz}$ and $d_{yz}$ orbitals, which are nearly empty.
For more negative $\Delta z$ (i.e., thinner slabs), the $d_{xy}$
orbital approach even more half-filling, while the $d_{xz}$ and
$d_{yz}$ orbitals are progressively emptied 
(see \fref{fig:svo-toten}, bottom). Additionally, the bandwidth decreases, enhancing thus the correlations and 
the size of the Mott gap (not shown here).
The main result of this total-energy calculation is that the minimum
calculated with DFT+DMFT is shifted significantly towards lower $\Delta z$
(\Fref{fig:svo-toten}, top),  
indicating that the structure with the lowest energy has a smaller slab width than obtained by DFT.

The trend in the structural change can be roughly explained in
  terms of an additional
energy gain by removing the degeneracy between the in-plane $d_{xy}$
and out-of-plane $d_{xz}$, $d_{yz}$ orbitals for smaller values of $\Delta z$. 
Indeed, having both types of orbitals occupied results in an additional energy cost proportional
to interorbital coupling $U - 3J$. Once the apical oxygen is moved sufficiently
close to the V ion, the anti-bonding orbitals $d_{xz}$, $d_{yz}$ are pushed up 
in energy and only $d_{xy}$ remains occupied (half-filled). This
removes the interorbital energy cost, lowering thus the total
energy.


\section{Conclusion}
\label{sec:conclustion}

We have presented a charge self-consistent implementation to combine
DFT with many-body techniques in the
\VASP{} package, based on optimized projector localized orbitals (PLO)
in the PAW framework. The implemented optimization, seeking the 
partial wave with the largest overlap with the relevant correlated subspace,  is crucial for
concise projections and leads to a straight-forward
connection between delocalized Kohn-Sham states and localized basis
functions. As usual,  in the localized subspace,  Hubbard-like
Hamiltonians can be used straightforwardly. In
contrast to a maximally-localized Wannier projection, the projector
formalism is very simple, easy to implement, preserves symmetry, and does not require any
special precautions for strongly entangled bands. Therefore, the projector formalism is also well suited for the simulation of correlation effects in supercells with a large number of bands. We have exemplified the benefits of using optimized projectors for the case of NiO.

We have benchmarked our fcsc implementation for two cases. First, we have compared a
standard \VASP{} DFT+U calculation with a charge self-consistent
mean-field treatment of the DFT+DMFT Hamiltonian for the case of NiO. We
find only small deviations for  the DOS, which we relate to different projections used in the standard \VASP{} DFT+U and the 
DFT+DMFT scheme. For NiO an important finding is that the double-counting
problem is alleviated by the charge self-consistency. With charge self-consistency
the influence of the double-counting parameter on the band gap is reduced by  
about a factor 2 compared to one-shot calculations.  

Second, we simulated a
SrVO$_3$ monolayer and found strong orbital polarization, which is
decreased in charge self-consistency. This  agrees with previously published
results using FLAPW+DMFT. Additionally, as a proof of concept, we
calculated the total energies from DFT+DMFT and found that correlation
effects lead to structural changes in SrVO$_3$ monolayers, reducing
the apical oxygen height in the single layer.

The presented projector and fcsc scheme can be used to interface
basically any many-body method with the \VASP{} package. It offers a robust and
concise interface for materials studies as well as future developments
of tools for strongly-correlated electron systems.

\ack

We thank Olivier Parcollet and Michel Ferrero for discussions about the \triqs{} interface.
Funding by the Austrian Science Fund (FWF) within  Project No.\ Y746 and
within the F41 (SFB ViCoM) is gratefully acknowledged.
O.~P. is grateful to A.~Georges for discussions on the details of charge
self-consistency.
O.~P. acknowledges support from the Swiss National Science Foundation NCCR MARVEL and 
computing resources provided by the Swiss National Supercomputing Centre (CSCS) under 
projects s575 and mr17.
The support from the Austrian federal government
(in particular from Bundes\-ministerium f\"ur Verkehr, Innovation und
Technologie and Bundes\-ministerium f\"ur Wirtschaft, Familie und Jugend)
represented by \"Oster\-reichische For\-schungs\-f\"orderungs\-gesellschaft mbH
and the Styrian and the Tyrolean provincial government, represented
by Steirische Wirt\-schafts\-f\"or\-der\-ungs\-ge\-sell\-schaft mbH and Standort\-agentur
Tirol, within the framework of the COMET Funding Programme is
also gratefully acknowledged. The work at the University of Bremen has been supported by the Deutsche Forschungs\-gemeinschaft (DFG) via FOR 1346 as well as the Zentrale Forschungs\-förderung of the University of Bremen.

\section*{References}
\bibliographystyle{iopart-num}
\bibliography{vasp_csc_bib}

\end{document}